\documentclass{ieeetran}
\usepackage{cite}
\usepackage{amsmath,amssymb,amsfonts}
\usepackage{algorithmic}
\usepackage{graphicx}
\usepackage{textcomp}
\usepackage{multirow}
\usepackage{soul}
\usepackage[caption=false,font=footnotesize,subrefformat=parens,labelformat=parens]{subfig}
\def\BibTeX{{\rm B\kern-.05em{\sc i\kern-.025em b}\kern-.08em
    T\kern-.1667em\lower.7ex\hbox{E}\kern-.125emX}}
\begin{document}

\title{QubiC: An open source FPGA-based control and measurement system for superconducting quantum information processors}
\author{\IEEEauthorblockN{Yilun Xu,$^1$
Gang Huang,$^1$
Jan Balewski,$^1$
Ravi Naik,$^2$
Alexis Morvan,$^1$
Bradley Mitchell,$^2$
Kasra Nowrouzi,$^1$
David I. Santiago,$^1$
and Irfan Siddiqi$^{1,2}$}
\IEEEauthorblockA{\\$^1$Lawrence Berkeley National Laboratory, Berkeley, CA 94720, USA
\\$^2$University of California at Berkeley, Berkeley, CA 94720, USA
\\Corresponding author: Gang Huang (email: ghuang@lbl.gov)}}

\maketitle

\begin{abstract}
As quantum information processors grow in quantum bit (qubit) count and functionality, the control and measurement system becomes a limiting factor to large scale extensibility. 
To tackle this challenge and keep pace with rapidly evolving classical control requirements, full control stack access is essential to system level optimization. 
We design a modular FPGA (field-programmable gate array) based system called QubiC to control and measure a superconducting quantum processing unit. 
The system includes room temperature electronics hardware, FPGA gateware, and engineering software. 
A prototype hardware module is assembled from several commercial off-the-shelf evaluation boards and in-house developed circuit boards. 
Gateware and software are designed to implement basic qubit control and measurement protocols. 
System functionality and performance are demonstrated by performing qubit chip characterization, gate optimization, and randomized benchmarking sequences on a superconducting quantum processor operating at the Advanced Quantum Testbed at Lawrence Berkeley National Laboratory. 
The single-qubit and two-qubit process fidelities are measured to be 0.9980$\pm$0.0001 and 0.948$\pm$0.004 by randomized benchmarking. 
With fast circuit sequence loading capability, the QubiC performs randomized compiling experiments efficiently and improves the feasibility of executing more complex algorithms. 
\end{abstract}

\section{Introduction}
The quantum computer represents a paradigm shifting innovation for computing technology, spurring the development of new breakthroughs in science \cite{preskill2018quantum,google2020hartree}.
Superconducting-circuit-based quantum bits (qubits) are a leading platform for quantum information science, with recent demonstrations of quantum advantage \cite{arute2019quantum}.
Qubit control hardware generates and routes complex sequences of radio frequency (RF) signals from room temperature electronics to the quantum processor at cryogenic temperature. 
As the size and complexity of the quantum system increases, the cost-efficient and compact generation of such signals becomes a bottleneck limiting system extensibility in the near term noisy intermediate-scale quantum (NISQ) computing era \cite{vainsencher2019superconducting}.
Thus, hardware elements in a next-generation design should be carefully matched to the needs of both current and potential future superconducting multi-qubit processors, while maintaining the modularity available at the printed circuit board (PCB) design level. 
However, commercially available lab equipment such as arbitrary waveform generators (AWG) and data acquisition cards (DAQ) are designed for general purpose test and measurement applications, and typically cannot keep pace both in terms of footprint and cost as quantum system complexity increases \cite{ryan2017hardware}. 
A customized quantum engineering solution rooted in extensible primitives is needed for the quantum computing community.

The field-programmable gate array (FPGA) architecture allows for customized solutions capable of growing and evolving with the field \cite{salathe2018low}. 
Several FPGA frameworks have been developed for quantum control and measurement \cite{steffen2013deterministic,riste2013deterministic,bultink2016active,gebauer2020state}.
Nevertheless, current FPGA-based control systems are not fully open to the broader quantum community so it is hard to make a general toolbox for information scientific discovery \cite{chen2012multiplexed,ofek2016extending,fu2017experimental}. 
Understanding the full electronics hardware and software stack of the qubit control systems is essential to the system level optimization and extensibility \cite{bertels2020quantum}.
Furthermore, compiling near-term quantum algorithms directly down to the native hardware descriptions via pulse-level control will improve the reliability of program execution \cite{gokhale2020optimized}.
Recent developments have also explored highly integrated system-on-chip (SoC) solution and moved it to the cryogenic stage \cite{bardin201928,patra2020scalable,pauka2021cryogenic}.
As such efforts mature, they will benefit from a deeper understanding of the control requirements that emerge as system size increases, and will thus rely on broad conceptual explorations rooted in more flexible hardware platforms. 

Here we develop and test the QubiC (Qubit Control) system -- an open source FPGA based RF control system that integrates qubit pulse generation and quantum state measurement \cite{qubic2021repository}.
Leveraging state-of-the-art FPGA technology, QubiC provides fully parametric waveform generation, and allows researchers to access all the control layers.
This scalable and cost-effective system will be a potential open source toolbox for the quantum community.

\section{QubiC system}
We design an FPGA-based qubit control system called QubiC to integrate the execution of gate-based algorithms with native quantum hardware implementation.
QubiC is an open source system which can accommodate users and developers from different layers to enable efficient co-design.
As the starting point, we develop and test a prototype system of a single control unit to demonstrate basic functionality and evaluate performance. 

The room temperature electronics hardware, the FPGA gateware, and the engineering software are integrated in the QubiC prototype system so as to realize the desired RF pulses to control and measure qubits. 
This prototype system also provides a platform to explore and optimize real-time feedback control such as fast reset \cite{vijay2012stabilizing,riste2012feedback}, and error correction algorithms.

\subsection{Electronics hardware}
\begin{figure}[t!]
\centering
\includegraphics[width=1.0\linewidth]{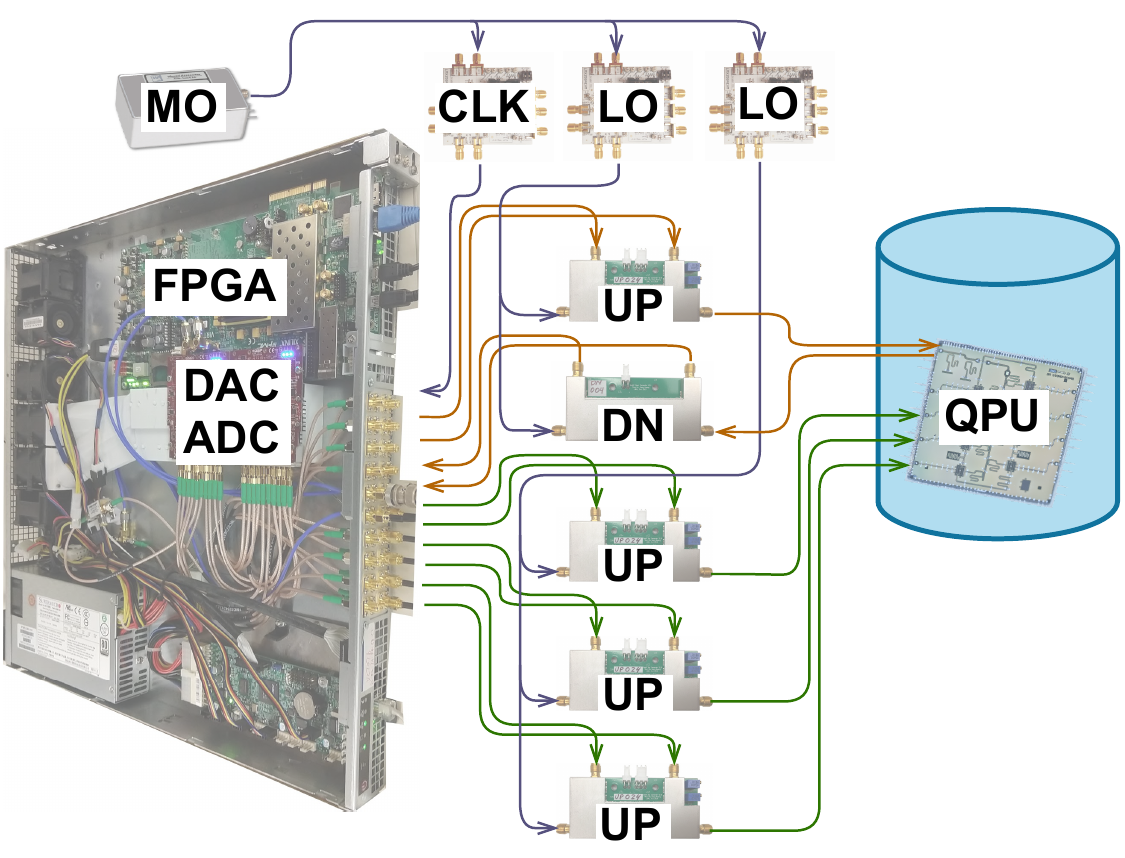}
\caption{QubiC prototype hardware. \textbf{MO}: master oscillator, \textbf{CLK}: clock, \textbf{LO}: local oscillator, \textbf{FPGA}: field-programmable gate array, \textbf{DAC}: digital-to-analog converter, \textbf{ADC}: analog-to-digital converter, \textbf{UP}: up converter, \textbf{DN}: down converter, \textbf{QPU}: quantum processor unit. The yellow line indicates the measurement path, while the green line defines the qubit control path.}
\label{fig:HW}
\end{figure}

The QubiC prototype hardware employs the heterodyne technique to generate and detect RF signals in a compact manner. 
It includes three basic building modules: the FPGA/ADC(analog-to-digital converter)/DAC(digital-to-analog converter) module to generate/detect the intermediate frequency (IF) signal; the RF mixing module to convert the signal frequency to/from the target frequency; and the local oscillator (LO) generation module to provide low noise LO signals, as shown in Fig.~\ref{fig:HW}.
A series of commercial off-the-shelf (COTS) evaluation boards are selected and assembled with in-house developed RF mixing modules to test functionality and benchmark performance.
Using these COTS evaluation boards enabled us to rapidly develop a hardware platform to study and develop the qubit control logic, while leaving plenty of room for channel density and cost optimization in future iterations.

\subsubsection{FPGA and ADC/DAC module}
The Xilinx VC707 and the Abaco Systems FMC120 boards are chosen for their computational capability and precision analog performance. 
The VC707 evaluation board contains an Xilinx Virtex-7 FPGA \cite{xilinx2012vc707}. 
Each FMC120 board has 4 channels of high speed (1.25~GSPS) 16-bit DACs and 4 channels of high speed (1~GSPS) 16-bit ADCs \cite{abaco2016fmc120}. 
One VC707 and two FMC120 boards are assembled in a 1U rack mount computer chassis with a customized cooling manifold to meet the FMC120 board cooling requirements. 
An external 1~GHz clock is provided to the chassis to run the DACs/ADCs at 1~GSPS, and to run the FPGA DSP at 250~MHz.
Compared with the latest RFSoC (radio frequency system-on-chip) chips, this combination gives us more choices with respect to DAC/ADC selection, particularly when tailoring hardware for specific future applications. 

\subsubsection{RF mixing module}
We develop compact RF up and down mixing modules that integrate the I/Q (in-phase/quadrature) mixer, IF/LO/RF power level adjustments and DC bias fine tuning on a 40~mm $\times$ 80~mm 4-layer PCB board with electromagnetic interference (EMI) shielding.
The RF mixing module is designed to work with RF and LO frequency between 2.5 and 8.5~GHz. 
Typical image rejection and adjacent channel isolation are measured to be $\sim$27~dBc and $\sim$50~dB. 
The RF mixing module provides 5$\times$10$^{-4}$ (V$_{\mathrm{pp}}$/V$_{\mathrm{mean}}$) amplitude stability and 1$\times$10$^{-3}$~radian (pk-pk) phase stability \cite{xu2020rf}. 

\subsubsection{LO generation module}
The noise performance of the LO generation module is critical for high-fidelity qubit operation because imperfections will directly map on to the RF/IF control signal through the RF mixing module. 
The LO frequency also needs to be adjustable from experiment to experiment to accommodate different chip designs. 
Additionally, some qubit operations will require multiple, different LO frequencies synchronized to each other.
At the same time, the size and cost of LO generation should also be factored into design consideration.

We thus use multiple phase locked loops (PLL) with a shared master oscillator (MO) to generate the LO and clock to meet these frequency and low-noise operation requirements.
The Wenzel (501-16843) 100 MHz ultra low noise crystal oscillator is used as the MO for the prototype system \cite{wenzel2006uln}.
The Texas Instruments LMX2595 evaluation modules are used as a PLL given its 20~GHz wide frequency range and low-noise performance \cite{ti2017lmx2595}.
Specifically, we measure the RMS jitter of the ``Wenzel+LMX2595'' module at the LO frequency at the qubit readout resonator frequency (6.52~GHz) and qubit drive frequency (5.50~GHz). 
As shown in Table~\ref{tab:LO}, the RMS jitters of ``Wenzel+LMX2595'' module are 1.0~ps (1~Hz--10~MHz) and 60~fs (100~Hz--10~MHz), which are comparable with the widely-used commercial Keysight signal generator \cite{keysight2021n5183b}.
The relationship between the LO phase noise metrics and the corresponding qubit dephasing spectral densities has been studied and the formula to estimate the operational fidelity bounds on superconducting qubits was derived in \cite{ball2016role}. 
Following the analysis and using the measured phase noise of the ``Wenzel+LMX2595'' configuration, we estimate that for the $\hat{X}$ gate shorter than 10~ms, the LO phase noise contributes $<$ $10^{-5}$ to the gate infidelity, which is much lower than the typical gate infidelity in superconducting qubits. 
This implies that the LO module phase noise is not the dominating factor to the gate fidelity.
Moreover, the ``Wenzel+LMX2595'' module is much more compact in size and lower in cost.

\begin{table}
\centering
\caption{LO RMS jitter}
\label{table}
\setlength{\tabcolsep}{3pt}
\begin{tabular}{c|c|c}
\hline
\multirow{3}{*}{Modules}&\multicolumn{2}{c}{RMS Jitter (@6.52~GHz / @5.50~GHz)}\\
\cline{2-3}
& Integration & Integration \\
& Bandwidth & Bandwidth \\
& (1~Hz--10~MHz) & (100~Hz--10~MHz) \\
\hline
Wenzel+LMX2595 & 1.0~ps / 0.9~ps & 60~fs / 54~fs \\
\hline
Keysight N5183B & 1.0~ps / 0.9~ps & 31~fs / 41~fs \\ 
\hline
\end{tabular}
\label{tab:LO}
\end{table}

\subsubsection{Module synchronization}
To scale the system up, we design the module-to-module synchronization using three layers of protocols \cite{huang2021clock}. 
\begin{itemize}
\item JESD204B subclass 1 \cite{ti2016jesd204b} is used to synchronize multiple DAC chips via the external SYSREF signal. 
\item Direct GPIO (general-purpose input/output) triggers are used to pass and fan out the event information which requires low latency among modules. 
\item Fiber-based synchronization using a protocol similar to the white rabbit \cite{moreira2009white} system is under development to lock the clock phases among the modules and also provide a high speed data communication path. 
\end{itemize}

\subsection{FPGA gateware}
The gateware programmed on the FPGA is the lowest level to implement digital tuneup sequences and algorithmic protocols. 
Modularity is the key to designing reusable and maintainable FPGA gateware code.
The QubiC FPGA gateware is written in the Verilog programming language and it is separated into three modules: the board support package (BSP), the digital signal process (DSP), and the host interface (HOI). 
\begin{itemize}
\item The \textbf{BSP} is hardware specific and used to implement low-level hardware configuration and initialization. 
\item The \textbf{DSP} implements basic qubit control and measurement functions and it is designed to be independent of the low-level FPGA/ADC/DAC selection. 
The DSP runs in a single clock domain and the HOI and BSP handle the clock domain crossing for the required registers, and also buffers the transmitted and received data.
\item The \textbf{HOI} handles all the input and output to and from the host computer, which is the interface that hosts higher level software.
\end{itemize}

The QubiC gateware is capable of generating waveform form parameters, which leads to the benefit of pulse sequence reuse.
The QubiC FPGA DSP is flexible and extensible.
In particular, the QubiC system has the capability of the fast reset by implementing the qubit status classification and the feedback loop in the FPGA gateware with low latency \cite{huang2020qubit}. 
In this section we will focus on the gateware DSP module and the host interface associated with it.

\subsubsection{Gateware DSP}
\begin{figure}[!ht]
\centering
\includegraphics[width=1.0\linewidth]{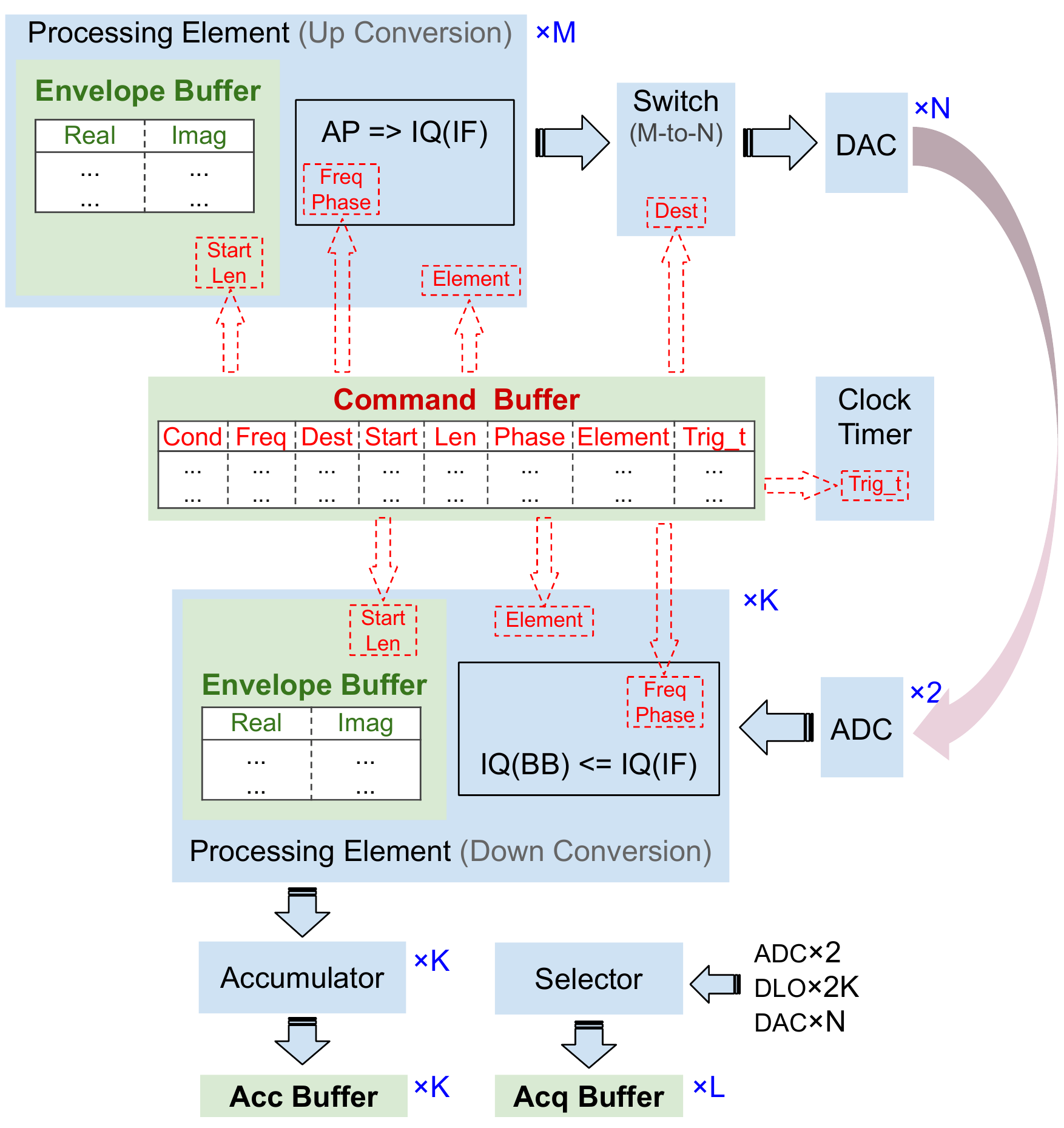}
\caption{QubiC gateware DSP block diagram. The \textbf{processing elements} are employed as the up or down converters in the digital domain. \textbf{Real} and \textbf{Imag} denote the real and imaginary parts of the complex envelope value in the \textbf{envelope buffer}. \textbf{AP} represents the amplitude (the magnitude of the complex envelope) and the phase (computed by the carrier frequency and initial phase from the command). \textbf{IQ(IF)} and \textbf{IQ(BB)} are the in-phase and quadrature-phase components of IF and baseband signals, respectively. The fields in the \textbf{command buffer} are condition (fast reset flag), frequency, destination, start, length, phase, element, trig\_t from the most significant bit (MSB) to the least significant bit (LSB). The accumulated values that comprise the integration of the baseband I/Q series over the corresponding DLO are stored in the \textbf{acc buffer}. The \textbf{acq buffer} serves as a live oscilloscope for the ADC/DLO/DAC raw data. M up conversion \textbf{processing elements} are switched to N \textbf{DACs} with the dynamic mapping. K down conversion \textbf{processing elements} can be assigned to process K-qubit readout data simultaneously through the same \textbf{ADC} pair.}
\label{fig:GW}
\end{figure}

The basic requirements of qubit control and measurement are to generate pulses at a specified carrier frequency with arbitrary amplitude modulation, and to also synchronously detect the pulse after it goes through the readout resonator.
The hardware module shifts frequencies between the RF and IF, while the gateware DSP module is responsible for the generation and processing of the data stream between the IF signal and the baseband signal.
The QubiC gateware DSP block diagram is illustrated in Fig.~\ref{fig:GW}.

The digital modulation/demodulation between the baseband envelope and the IF data stream is realized by a module called the processing element.
Processing elements can be configured to execute digital up or down conversions.
When used as the up converter, the module generates IF pulses in-phase ($I_{\mathrm{IF}}$) and quadrature-phase ($Q_{\mathrm{IF}}$) components from the gate-specific baseband envelope $I_{\mathrm{BB}}+jQ_{\mathrm{BB}}$, the initial phase $\phi_0$ and the digital local oscillator (DLO) frequency, as in
\begin{equation}
\label{eq1}
I_{\mathrm{IF}} + jQ_{\mathrm{IF}} = (I_{\mathrm{BB}} + jQ_{\mathrm{BB}}) e^{\phi_0} e^{j\omega_{\mathrm{DLO}}t} .
\end{equation}
The carrier frequency is determined by the physical property of the qubits or the readout resonator.
The carrier phase naturally rotates as time evolves; additionally the engineering software can inject additional phase offsets to implement virtual Z gates \cite{mckay2017efficient}.
The pulse envelope is defined or optimized through the qubit gate calibration process, and stored in the FPGA memory point-by-point as complex numbers, each with in-phase and quadrature-phase terms. 
Moreover, this pulse envelope can be used repetitively by indexing the same address area, even if the carrier frequency or phase changes.

The calculated result then feeds into any DAC through an m-to-n switch.
Signals from multiple processing elements can be sent to the same DAC and added together.
For each pulse to be generated, the software needs to specify the processing element, and the DAC to be used for the destination.

When the processing module is used in the down conversion mode, it receives the qubit signals from the ADC, and recovers the baseband I and Q components, which is exactly the opposite operation, as defined by
\begin{equation}
\label{eq2}
I_{\mathrm{BB}} + jQ_{\mathrm{BB}} = \left( I_{\mathrm{IF}} + jQ_{\mathrm{IF}} \right) e^{-j\omega_{\mathrm{DLO}}t} .
\end{equation}
Then the baseband I/Q series are integrated by a vector accumulator, and the results are stored in an ``accumulation buffer'' (acc buffer) for the corresponding channel to determine the qubit final state. 
The raw ADC, DAC, DLO values are available to the host computer through ``acquisition buffers'' (acq buffer) and the integrated I/Q data are available to the host computer through the acc buffer.

In the QubiC gateware, the processing element is flexible in the waveform generation.
Each processing element has a waveform storage and a phase rotator, but only puts out one waveform at a time. 
Once stopped, it is available to replay another waveform with another configuration. 
Waveform envelopes drawn from that memory can be variable lengths in order to start and stop arbitrarily.
Currently, the envelope buffer of each processing element is 1k deep and 32 bits wide. 
The upper and lower 16 bit words of each envelope point represent the real (I) and imaginary (Q) parts of the pulse envelope respectively.
The conversions between amplitude/phase and I/Q of a complex signal as in \eqref{eq1} and \eqref{eq2}, are implemented with the CORDIC (COordinate Rotation DIgital Computer) algorithm on FPGAs \cite{volder1959the,andraka1998a}.

\subsubsection{Host interface}
A qubit algorithm typically consists of a series of qubit gates and measurements, which are eventually realized by the sequence of RF pulses.
In general, a gate may consist of a series of pulses.
Gate level experiments adopt a pre-calibrated gate pulse, while pulse level experiments need access the parameters for each pulse. 
The host computer interface needs to be designed to accommodate both types of experiments by way of parametric waveform generation.

In the current QubiC gateware, each pulse is defined by the pulse start time, the envelope information, and the carrier information. 
We define a 128-bit command as the interface between the lower level gateware on an FPGA and the higher level software on a host computer:
\begin{itemize}
\item \textbf{Trig\_t} (24 bits) Pulse start time relative to the start of the whole sequence.
\item \textbf{Start} (12 bits) The start address of the envelope data in the envelope buffer.
\item \textbf{Length} (12 bits) The length of envelope buffer to play for this pulse.
\item \textbf{Frequency} (24 bits) The carrier frequency in 1~GHz represented by a 24 bits word. Each step is 1~$\mathrm{GHz}/2^{24}\approx60~\mathrm{Hz}$.
\item \textbf{Phase} (14 bits) The carrier initial phase as the $\phi_0$ shown in (\ref{eq1}). Each step is $360^\circ/2^{14}\approx 0.022^\circ$.
\item \textbf{Element} (8 bits) Processing element index.
\item \textbf{Destination} (2 bits) Destination IQ pair index.
\item \textbf{Condition} (1 bit) Flag for conditional gate, used for the fast reset gate (only available on the specified version of bit file).
\item \textbf{Reserved} 31 bits
\end{itemize}

The 128-bit command provides the parametric waveform generation from the lowest FPGA level.
The QubiC reuses sequences by updating waveform parameters via commands without recompiling or resending the waveform data.

The commands defining quantum circuits are stored in a 64k deep 128-bit wide buffer on an FPGA.
The envelope buffers are also written from the host computer and stored in each processing element. 
All commands repeat at a defined period until the acc buffer is full. 
The acc buffer can be cleared by the software after reading the data.


Writing to the command/envelope buffer to generate pulses, and reading from the acc/acq buffer to collect responses are the lowest level software interface provided by the FPGA gateware.
At this level, we have to adhere to the restrictions imposed by the gateware implementation, for example, the maximum pulse envelope length and command depth, the frequency and phase resolution, the longest sequence time etc.
The complexity of the restrictions imposed at this lower level makes it more difficult to use directly. 
However, the advantage is that this layer has full access to the FPGA directly, and mainly serves developers who implement software directly on top of the QubiC gateware. 

\subsection{Engineering software}
The engineering software runs on the host computer, implementing the FPGA buffers/registers input/output and providing the software interface to higher level software or algorithms. 
QubiC engineering software consists of an application programming interface (API), a portfolio of quantum characterization verification and validation (QCVV) experimental scripts and a graphical user interface (GUI). 

\subsubsection{Application programming interface}
The Python 3 based API compiles the quantum processor gate pulse specification and the quantum circuit description to gateware commands and pulse envelopes.

The quantum processor gate pulse specification contains two parts: the frequencies (readout resonator and qubit drive) that represent the specific properties of each qubit and the gates that drive interactions between the control system and the qubits. 
Each gate is a series of pulses defined by its destination, timing, carrier, amplitude and envelope.
The quantum processor and gate specifications are captured in the JSON format file, which is generated or updated during the chip calibration.

The quantum circuit description can be written in our native function call or be imported from TrueQ \cite{trueq2020}. 
The measurement result can also be sent back to TrueQ to allow for post-processing of all the data. 
A similar interface to Qiskit/OpenQASM/Cirq \cite{qiskit2020,cross2017open,cirq2020} is under development.

The compiler is the bridge to translate the circuit description into the envelope and command buffers.
There are two distinct types of compilers in the QubiC API. 
The first one is the OPTM (OPTiMize gate) compiler which is used during the chip characterization and gate optimization.
The OPTM compiler accommodates multiple different pulse envelopes, in length, shape or other parameters.
The envelope buffers from all the processing elements are collected together and allocated dynamically so that they can all be used in the specified qubits. 
In the OPTM compiler, the circuit description is written in our native function call.
One can add a gate onto a sequence at a specific time and modify the gate parameters at the compiling time.
The validation test is necessary in the OPTM compiler and so it takes longer to compile.

Another compiler is the RUNC (RUN Circuit) compiler, which is used to run a quantum circuit with calibrated gates on all the qubits simultaneously.
Since only a few predefined gates are applied for each qubit, the envelope memory can be pre-allocated statically to reduce the compiling time. 
In the RUNC compiler, the circuit description can be imported from other languages, which need to be translated to the QubiC format. 
The validation tests during compiling are skipped to execute long circuits efficiently in the RUNC compiler.


The OPTM and RUNC compilers coexisting in the QubiC software bring in the flexibility and efficiency for distinct control flows. 
The OPTM compiler targets a comprehensive chip calibration via tweaking multiple parameters with high accuracy in a broad range, while the RUNC compiler specifies a dedicated structure to perform the advanced quantum algorithms in an efficient manner.
Both the OPTM and RUNC are 3-step compilers which are illustrated in Appendix.

\subsubsection{QCVV experiments support}


A calibrated quantum information processor is the necessary prerequisite to perform QCVV experiments. 
In the current QubiC engineering software, a series of scripts are developed to characterize the quantum information processor and optimize the gates to support the superconducting QCVV experiments.
This type of characterization and optimization routine can also be adapted for other architectures at hand.
Quantum processor characterization begins with time alignment, which measures the latency for the readout signal and aligns it with the DLO.
The single-tone experiment involves measuring all the readout resonator frequencies on the readout bus \cite{chen2012multiplexed}.
We also conduct a punch-out experiment where the drive power is increased significantly to suppress the Josephson nonlinearity of the qubit, yielding a simple spectroscopic signature of viable qubits, and a coarse adjust of the readout amplitude needed to achieve linear operation \cite{reed2010high}.
The qubit drive frequency can be extracted from a two-tone spectroscopy experiment or the chevron pattern obtained from coherent oscillations.
We drive Rabi oscillations with different pulse lengths or amplitudes, and derive a Gaussian mixture model (GMM) for qubit state discrimination. 
A readout correction is employed to statistically calibrate out the effect of classical readout bit-flip errors.

A bounded minimization method \cite{brent2013algorithms} is used to adjust the readout amplitude and frequency together with the qubit drive frequency to maximize the Rabi oscillation contrast. 
To increase the gate repeatability, the gate pulse length is set to be a multiple of the DAC sampling rate and the FPGA clock rate. 
The qubit drive amplitude is optimized for each gate by stacking multiple identical gates at the target pulse width to maximize the recovery probability \cite{sheldon2016characterizing,xu2020automatic}.
With the calibrated single-qubit gate, we can measure the qubit coherence time including the qubit relaxation time ($T_1$) and dephasing time ($T_2$).
The residual oscillation frequency measured by the Ramsey experiment can help us finely tune the qubit drive frequency.
The single qubit gate quality can be validated by the AllXY experiment \cite{reed2013entanglement} and by a conventional randomized benchmarking (RB) sequence \cite{knill2008randomized}. 

Two-qubit gate calibration starts with optimizing the cross resonance (CR) pulse to reach full entanglement \cite{sheldon2016procedure}.
The single qubit gate parameters around the CR pulse to construct a CNOT gate are identified by a curve fitting of a full XY-plane measurement of a degenerated state followed by a CR pulse \cite{xu2021automatic}. 
With the circuits from TrueQ, we can execute a two-qubit RB measurement to obtain the two-qubit process fidelity.

\subsubsection{Graphical user interface}
The QubiC GUI (graphical user interface) is being developed to streamline and simplify the complex sequences of hardware calibration tasks.

The GUI client, which is implemented in Bootstrap and JavaScript,  can be used to remotely control the quantum hardware connected by the internet.

\begin{figure}[!ht]
\centering
\includegraphics[width=1.0\linewidth]{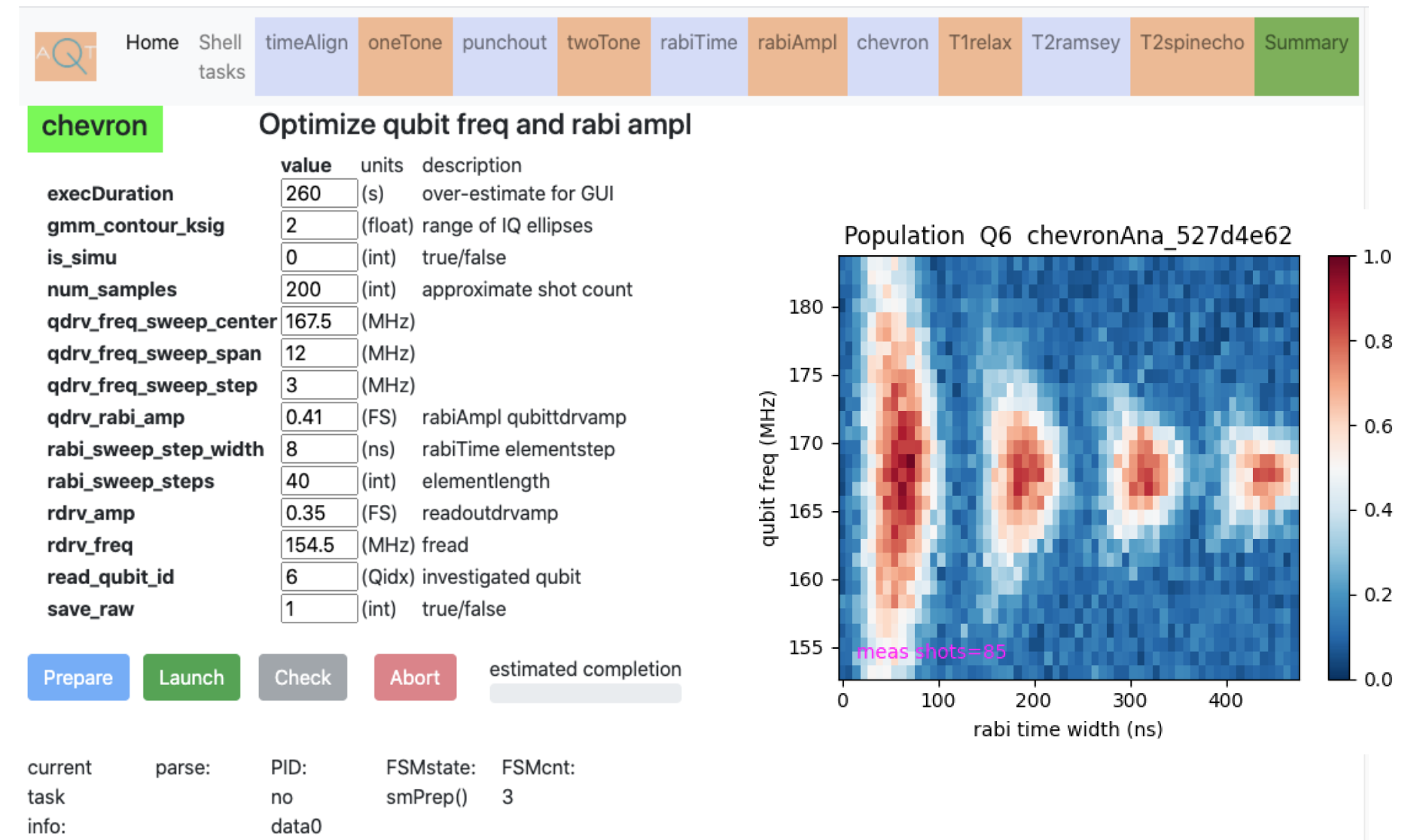} 
\caption{Screen shot of GUI while executing the chevron task.}
\label{fig:GUI}
\end{figure}

The operator can change the qubit bias point in real time, initialize/terminate measurements, and display the results. 
Fig.~\ref{fig:GUI} shows an example of the user-facing GUI web-page for generating a chevron pattern for coherent oscillations. 

The GUI server, which is implemented in the Python web framework Flask, runs on a Linux machine located in the proximity of the QubiC FPGA chassis. 
For any measurement requested by the client, the configuration and the measured raw IQ-pairs are saved in local files in the YAML and HDF5 data formats, respectively.
The post processing is done on the server side and the plots are made available for the remote client. 
Any file saved by the GUI server is time-stamped, which allows for the retrieval of the history or tracking of the time stability of the hardware calibration constants.
The finite state machine governs the internal logic of the server and the experiments are executed asynchronously allowing for an early termination by the client.

\section{Bench test}
Putting the electronics hardware, FPGA gateware and engineering software together, we used the QubiC system to generate the desired RF pulses to control and measure the qubits.
Specifically, we performed bench testing to demonstrate fully parametric waveform generation using the destination, timing, carrier, amplitude and envelope parameters.

\begin{figure}[t!]
\centering
\subfloat[Gate configuration and envelope definition. The destination parameter is the destination qubit or readout to be controlled, which is named as \textbf{dest} in the gate configuration. The timing parameter consists of the pulse starting time (\textbf{t0}) and the pulse width (\textbf{twidth}). The carrier frequency (\textbf{fcarrier}) and the initial phase (\textbf{pcarrier}) values are encoded in the carrier parameter. The pulse amplitude (\textbf{amp}) is a fraction of DAC full scale. The envelope function (\textbf{env}) can be any Python function (commonly used or customized) returning a NumPy array.]{\includegraphics[width=1.0\linewidth]{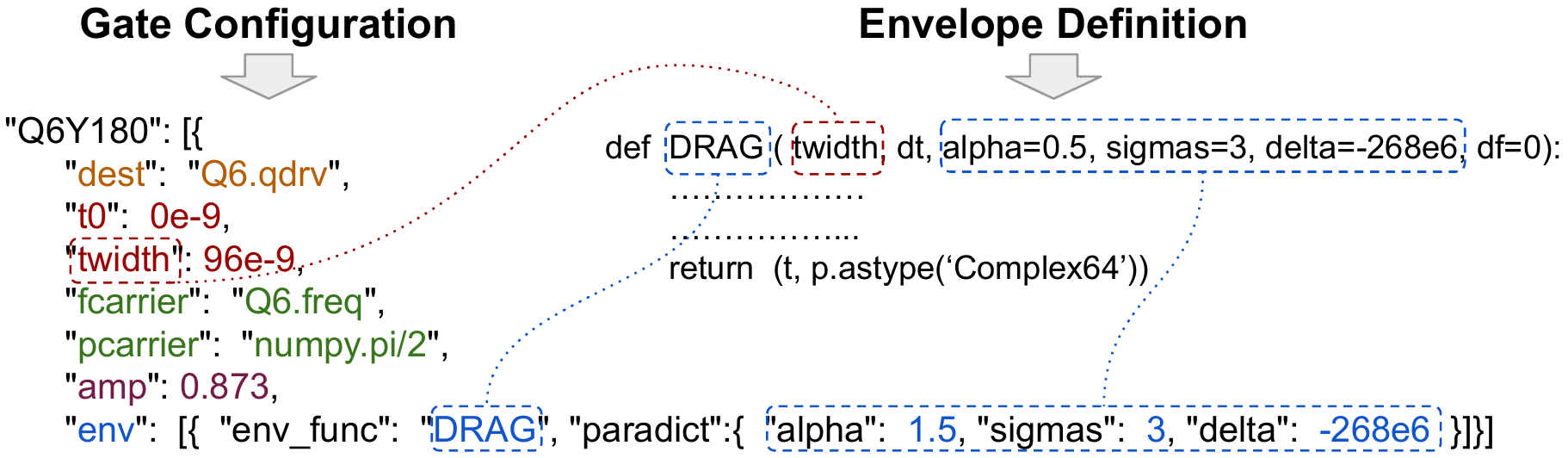}
\label{fig:bencha}}
\vfil
\subfloat[Y180 gate with DRAG envelope measured by ADCs. The ADC digitizes the IF signal at a sampling rate of 1~GSPS.]{\includegraphics[width=0.8\linewidth]{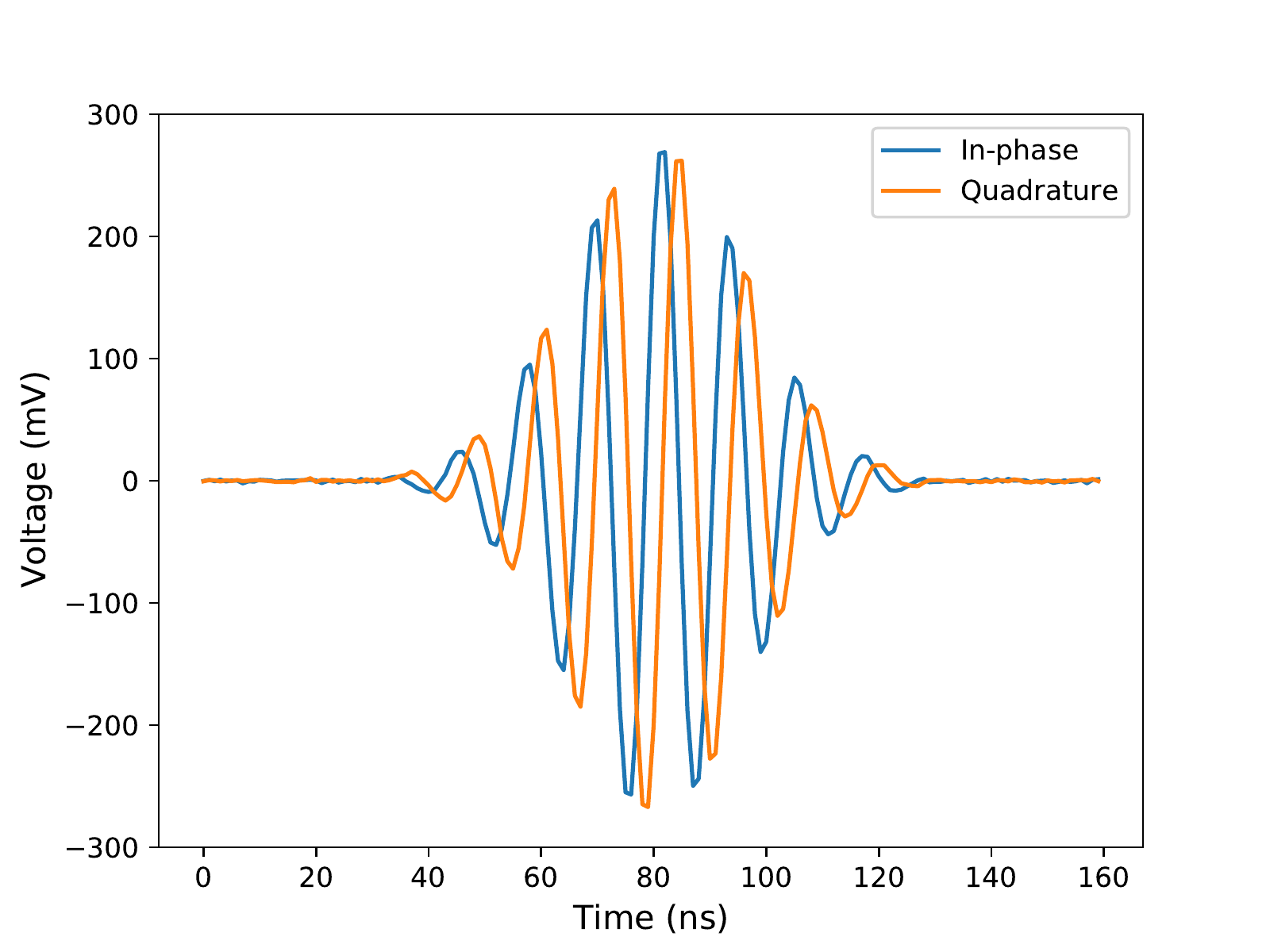}
\label{fig:benchb}}
\caption{Parametric waveform generation.}
\label{fig:bench}
\end{figure}

A simple example is shown in Fig.~\ref{fig:bencha} where we aim to generate a Y180 gate on Q6 (which is a 180 degrees single-qubit gate around the Y-axis of the Bloch sphere on a qubit labeled Q6) starting at some time (defined in the circuit) after the circuit starts.
The in-phase and quadrature components are sent to the IQ channels for the Q6 qubit drive ``Q6.qdrv'' (dest).
The pulse starts immediately (t0) after the gate starts, and lasts 96~ns (twidth).
The carrier frequency is specified as ``Q6.freq'' (fcarrier), while the initial phase is ``numpy.pi/2'' (pcarrier).
The overall amplitude of the pulse is 0.873 of the DAC full scale.
The amplitude modulation is applied with a ``DRAG'' (derivative removal by adiabatic gate \cite{motzoi2009simple}) envelope (env) with appropriate parameters.
This simple mapping from the gate configuration to the envelope definition function enables us to use any user provided Python function as the pulse envelope to construct a gate.

The above-mentioned Y180 gate pulse is generated by DAC pairs as IF I/Q signals. 
The IF signals are then up-converted in the up mixing module to an RF signal. 
In order to measure the Y180 gate pulse, the RF signal is directly connected back to the RF down mixing module, followed by the ADC digitization.
As shown in Fig.~\ref{fig:benchb}, we can successfully generate the parameterized RF pulse, which is essential to the pulse level experiments.


\section{Test with quantum processor}
\subsection{Randomized benchmarking}
\begin{figure}[t!]
\centering
\subfloat[Single-qubit RB. 100 points per length, 1k shots per point.]{\includegraphics[width=0.8\linewidth]{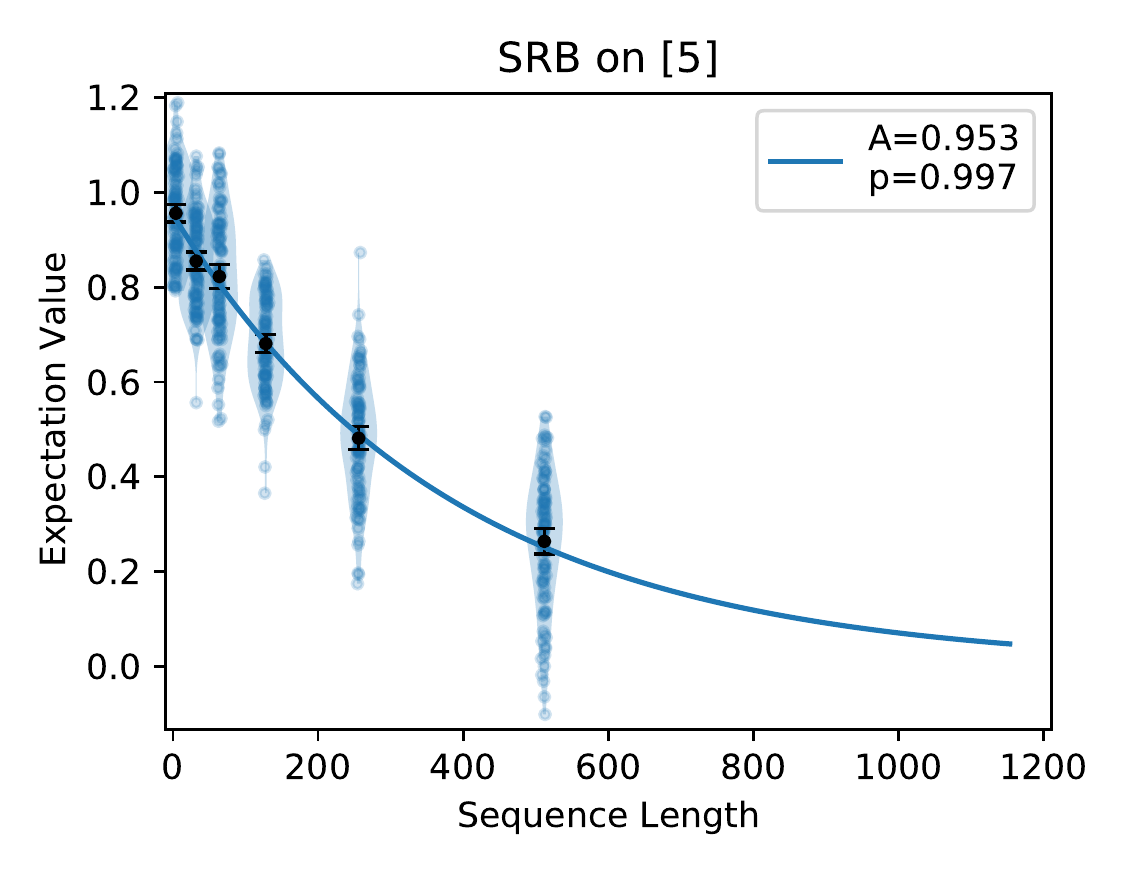}
\label{fig:rba}}
\vfil
\subfloat[Two-qubit RB. 20 points per length, 2k shots per point.]{\includegraphics[width=0.8\linewidth]{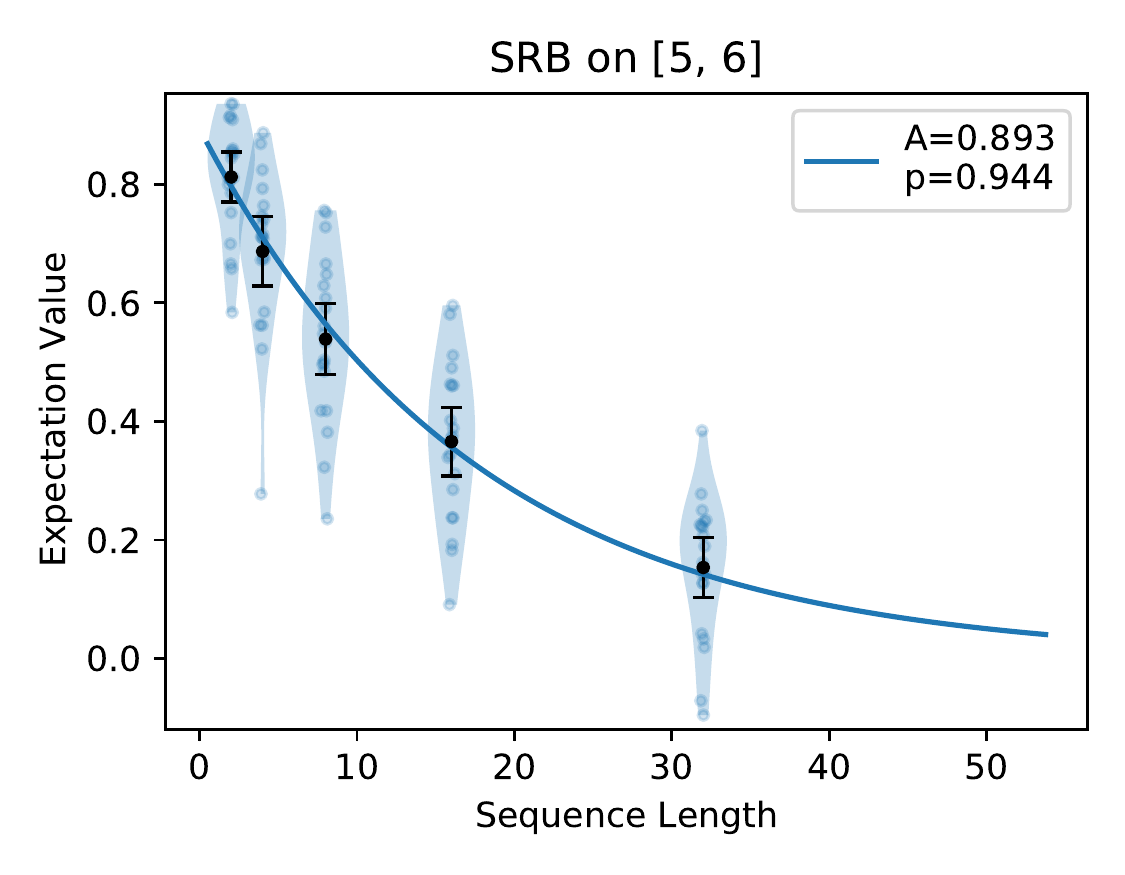}
\label{fig:rbb}}
\caption{Streamlined randomized benchmarking results. Each point corresponds to a measurement result of a random quantum circuit. The curve was fitted with an exponential decay function $\mathrm{A}\mathrm{p}^\mathrm{m}$. The sequence length is expressed in terms of Clifford gates.}
\label{fig:rb}
\end{figure}

Leveraging standard qubit characterization and gate optimization protocols \cite{xu2021automatic}, the QubiC can automatically find and tune a two-qubit system. 
In order to validate the performance of the QubiC system on quantum hardware, we perform streamlined randomized benchmarking \cite{magesan2011scalable} on two of the qubits in an 8-qubit quantum processor \cite{blok2021quantum}. 
As shown in Fig.~\ref{fig:rb}, the single-qubit process fidelity \cite{emerson2005scalable,dankert2009exact} is measured to be 0.9980$\pm$0.0001, while the two-qubit process fidelity is measured to be 0.948$\pm$0.004. 
The RB results demonstrate that the QubiC system can control and measure qubits efficiently, and it should be capable of delivering high-fidelity gates on state-of-the-art processors.

\subsection{Randomized compiling}
Randomized compiling (RC) is newly developed technique in quantum computing to tailor and mitigate noise \cite{wallman2016noise,ware2018experimental}.
The RC protocol efficiently compiles an algorithm circuit in a number of different, but equivalent variations. 
Instead of executing just the original ``bare'' circuit, the protocol requires execution of all the variations of the circuit. 
On the conventional control hardware side, it may be challenging to create and upload the various pulse sequences without significant overhead. 
Here, we show that QubiC can execute this general, yet sophisticated compilation protocol on a quantum computer in an efficient manner. 

On the host computer, the circuits are randomly compiled (``Compile'') and transpiled to a device's native gateset (``Transpile'') \cite{hashim2020randomized,trueq2020}.
Afterwards, QubiC transfers the native gate sequences to QubiC command sets (``Transfer'') and then generates the sequences (``SeqGen''). 
The QubiC software prepares the hardware and transports the command and the memory to the QubiC hardware (``Run''). 
The qubit response is acquired from the FPGA (``Acquire'') and processed by QubiC software to discriminate the state (``Process''). 

To evaluate RC efficacy, we compare the measured probability $\mathrm{P}(x)$ with the ideal probability $\mathrm{P}_\mathrm{ideal}(x)$ for the given circuit to calculate the total variational distance (TVD), which is defined as
\begin{equation}
\label{eq3}
\mathrm{TVD}=\frac{1}{2}\sum_{x \in X}|\mathrm{P}(x)-\mathrm{P}_\mathrm{ideal}(x)| ,
\end{equation}
where x is the bit string. 
The TVD parameterizes the probability of measuring an incorrect solution.

\begin{figure}[!ht]
\centering
\includegraphics[width=0.8\linewidth]{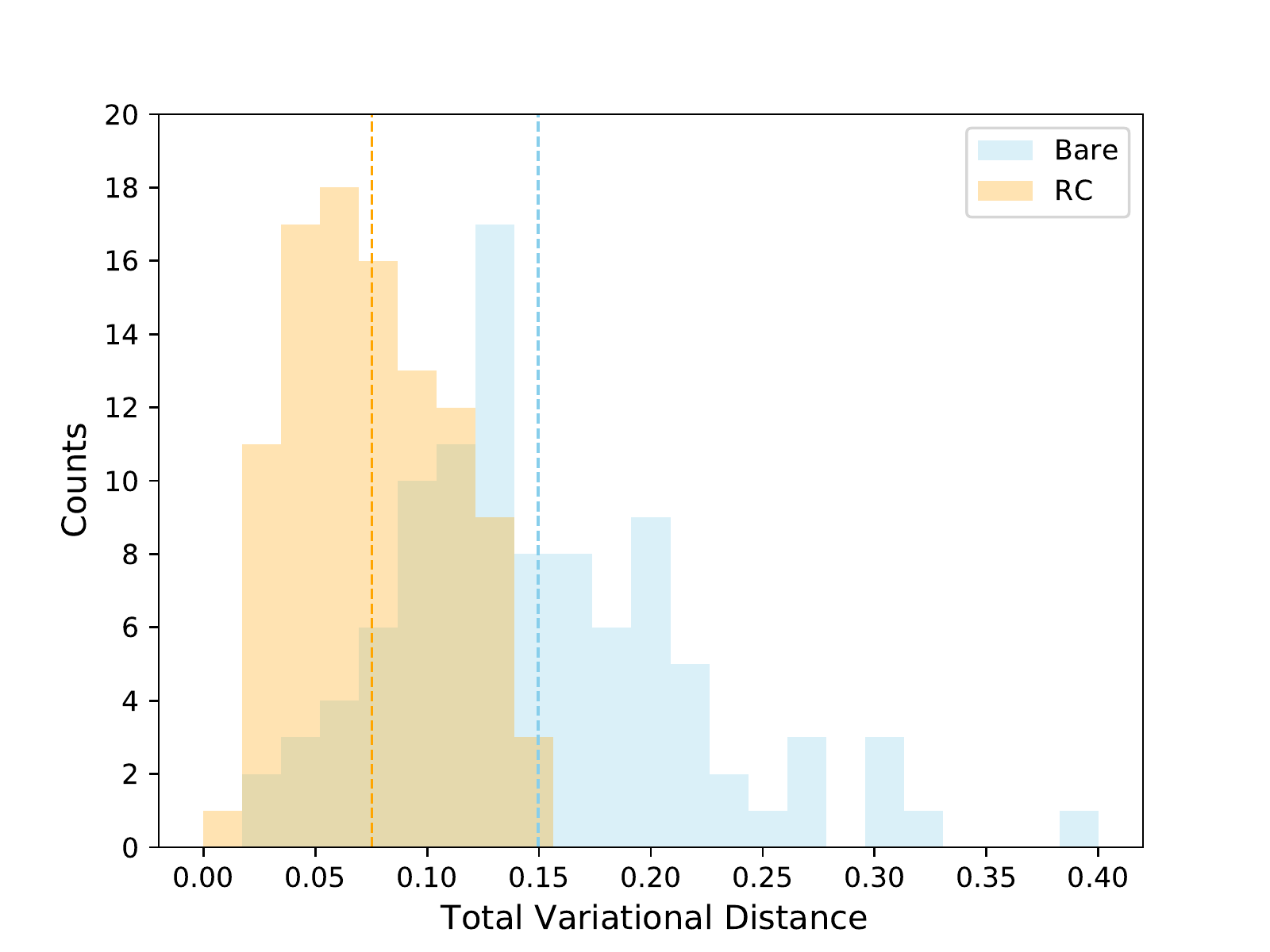} 
\caption{TVD distributions obtained from two-qubit randomized compiling. 100 different random bare circuits were generated for sufficient statistics, with an average of 20 RC circuits for each bare circuit. The circuit depth is 5 (two-qubit gates). Each measurement is the average of 1024 shots. The TVD values are 0.150$\pm$0.067 for bare circuits and 0.075$\pm$0.034 for RC circuits.}
\label{fig:rcdist}
\end{figure}

We perform the two-qubit RC on the quantum processor, and measure the TVD distributions of bare circuits and RC circuits, as shown in Fig.~\ref{fig:rcdist}. 
Compared to the bare circuits, lower TVD values are observed from the RC process, which means the output distributions are more accurate.

\begin{figure}[t!]
\centering
\subfloat[Execution time versus number of RC circuits.]{\includegraphics[width=0.8\linewidth]{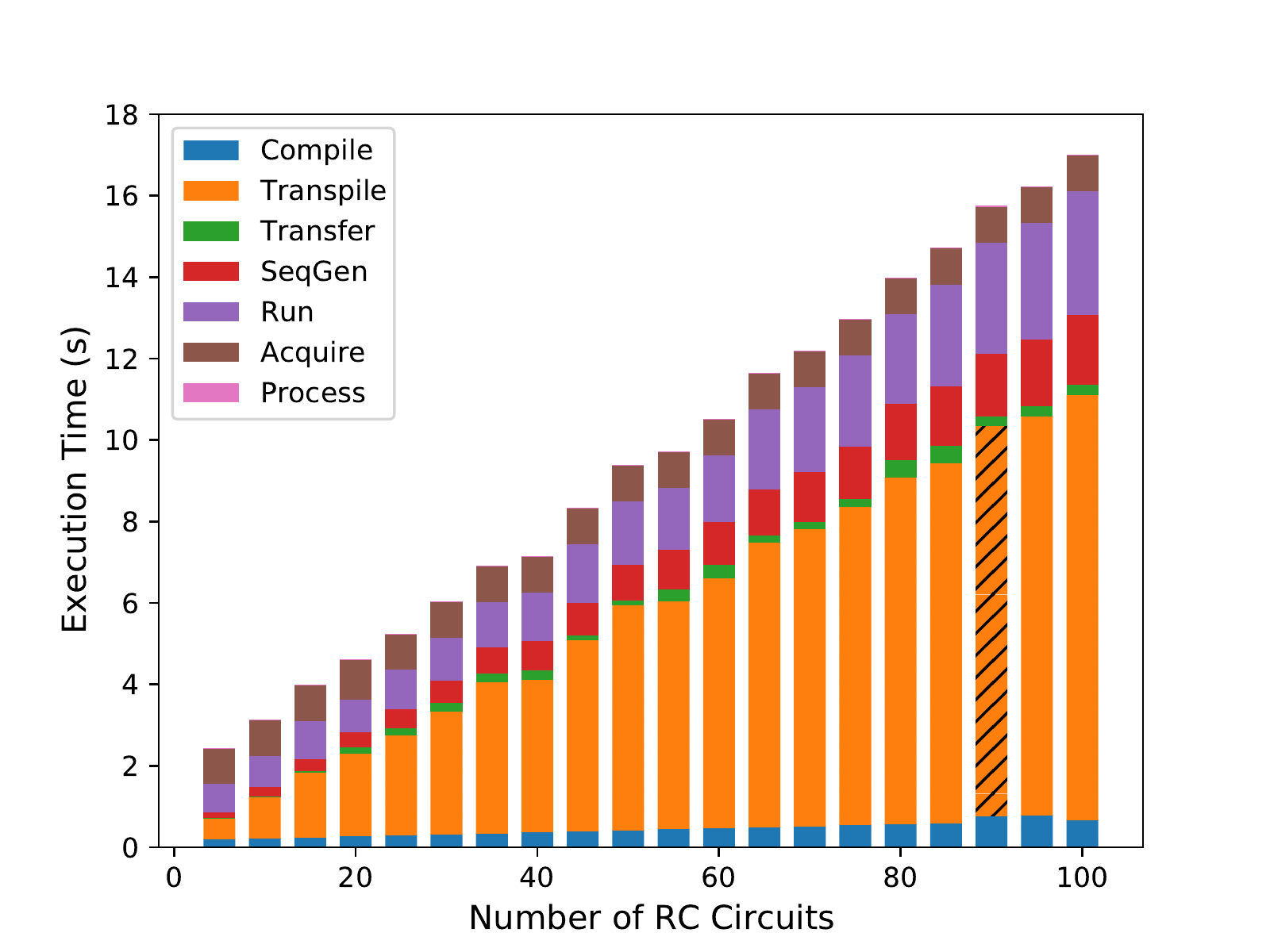}
\label{fig:rcspeeda}}
\vfil
\subfloat[Execution time versus number of two-qubit gates.]{\includegraphics[width=0.8\linewidth]{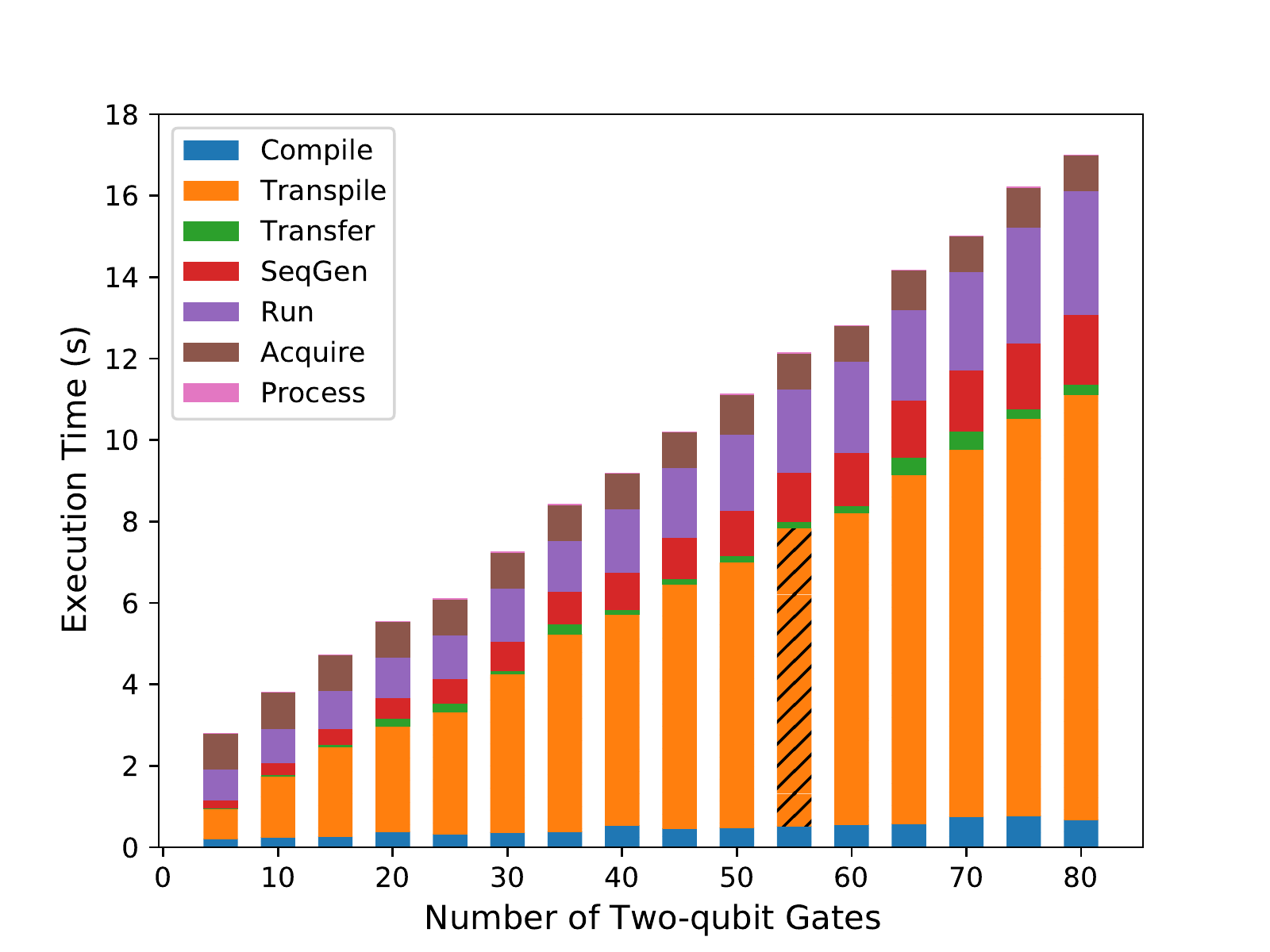}
\label{fig:rcspeedb}}
\caption{Execution time of two-qubit randomized compiling on QubiC. (a) measures the execution time with different number of RC circuits when the circuit depth is 80 (two-qubit gates). (b) measures execution time with different number of two-qubit gates when 100 RC circuits are averaged for each bare circuit. Each measurement is the average of 1k shots. The bar with slash inside is operated on the host computer, which is the main limiting factor in the execution time.}
\label{fig:rcspeed}
\end{figure}

Furthermore, QubiC can execute the RC protocol relatively quickly, as shown in Fig.~\ref{fig:rcspeed}.
The total execution time grows linearly with the increase of the number of RC circuits or the number of two-qubit gates.
The ``Compile'' and the ``Transpile'' account for the majority of the execution time, which are limited by the host computer performance.
The major part of the QubiC execution time is the ``Run'', while command writing is the most time-consuming task in the ``Run''. 
One can observe that the ``Run'' time increases linearly with the number of RC circuits and the number of two-qubit gates respectively, since the command depth also grows linearly. 
The speed of the ``Run'' portion is currently limited by the constraint that the load command only accepts 16k commands and takes $\sim$55~ms to execute. 
This can be made more efficient in future versions of the software.
For a 600~$\mu$s relaxation time, we can load $\sim$100 circuits at a time. 
Currently QubiC takes $<$17~s for 80 cycles and 100 RC circuits with 1024 shots, which is efficient and promising.

\section{Conclusion}
We develop an open source FPGA based QubiC system which bridges the quantum algorithm and the quantum hardware implementation layers of the computing stack.
The QubiC system includes the room temperature electronics hardware, the gateware, and the engineering software running on it to implement the qubit control pulses, and also to perform quantum state readout. 
QubiC allows researchers to access the full electronics hardware, gateware and software stack, which will enable the execution of a broader class of computation experiments while also facilitating the implementation co-design at each level of the stack in next generation systems. 
The QubiC prototype system functionality and performance were demonstrated by measuring single-qubit and two-qubit process fidelities. 
We obtained RB values of 0.9980$\pm$0.0001 and 0.948$\pm$0.004, respectively, on a 8-qubit superconducting quantum processor operating at the LBNL Advanced Quantum Testbed.
Moreover, we demonstrated that QubiC has the capability to run advanced quantum algorithms such as randomized compiling with promising execution speed. 
QubiC can accommodate the full spectrum of users and developers, and will be a potential open source toolbox for the quantum community.

\section*{Appendix}
\label{appendix}

\subsection{Compiling procedure}
The compiler is the bridge in the software API to transfer the circuit description into the envelope and command buffers.
The compiling procedure is shown in Fig.~\ref{fig:compiler}.
The three steps of the compilation are:

\begin{figure}[!ht]
\centering
\includegraphics[width=.6\linewidth]{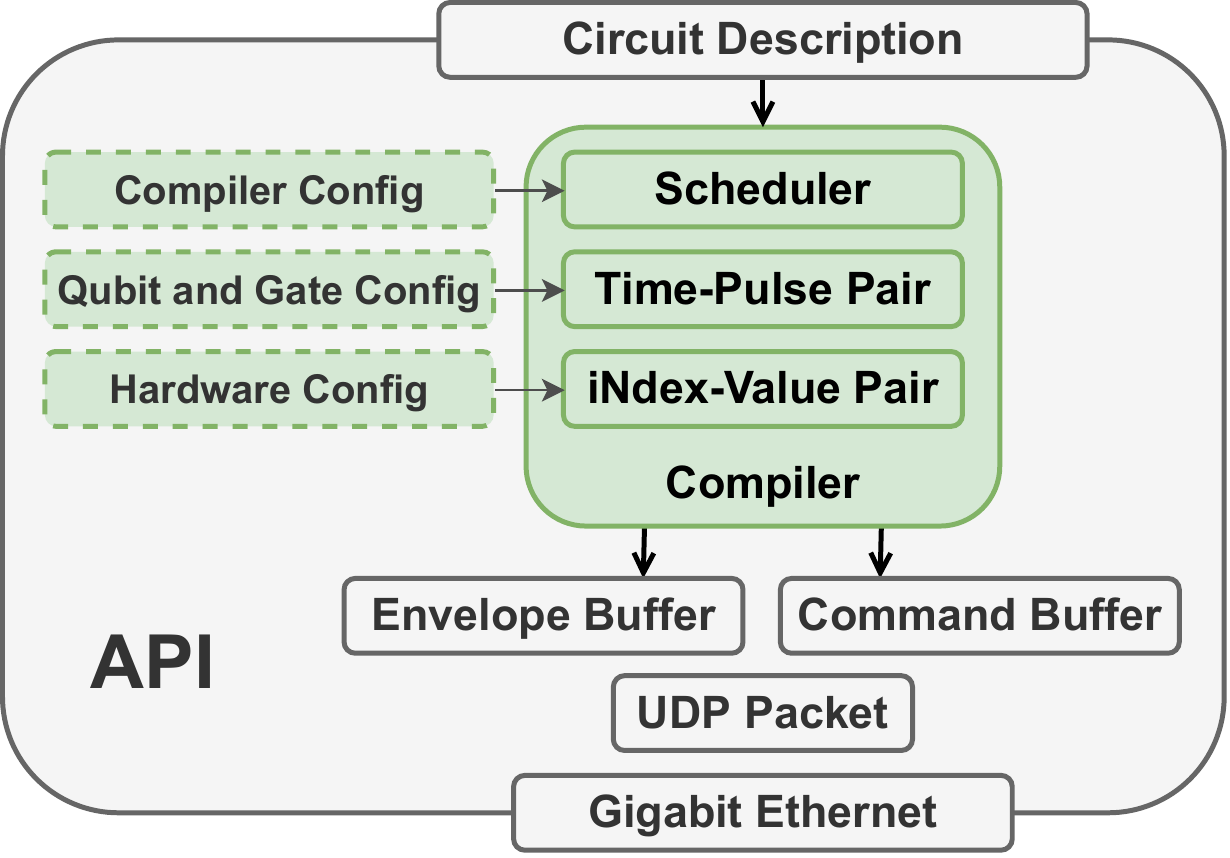} 
\caption{Compiling procedure.}
\label{fig:compiler}
\end{figure}

(i) \textbf{Scheduler}: 
The first step is scheduling, which takes in the circuit description together with the compiler configuration to generate the gate sequences with relative timing information. 
For the circuits imported from other language which does not include the timing information, such as OpenQASM, this step is necessary. 
However, if the circuit is written using native function calls, this step can be skipped by assigning an explicit start time to each gate.

(ii) \textbf{Time-Pulse Pair}: 
The second step is to combine the gate sequences with the gate configuration to generate a list of pulses described as TP (time pulse) pairs. 
The qubit chip and gate configuration are generated from the qubit characterization and gate optimization. 
The gate configuration contains information about the pulses used in the gate, including the carrier, the envelope, and the relative timing.
The TP description contains the parameterized gate information, which can then be implemented using different electronics hardware setups. 
The conventional AWG contains idle periods of zero or constant output. 
Rather than inefficiently storing repeated values in the waveform memory, we instead employ the TP pairs to store the time and pulse envelope value in the waveform memory.
At this level, the TP pair can still be human readable, so for the user who is familiar with the pulse should be able to understand the TP description and also check the relative pulse timing.
    
(iii) \textbf{iNdex-Value Pair}: 
The third step includes the hardware configuration, and generates the NV (iNdex value) pair as buffer value for the lower level. 
This is a hardware dependent layer. 
The hardware configuration contains the sampling rate, the physical wiring diagram between the DACs and the qubits and the gateware register maps. 
This only needs to be updated when the quantum chip or the chip wiring changes. 
Separating this step and configuration gives us the opportunity to run the same pulse on different hardware for comparison.

The outputs of the 3-step compiler are the envelope buffer, the command buffer and the simulator to visualize the pulse from the buffer values. 
The commands normally sent to the FPGA are redirected to the on-host analysis and visualisation package, when switching to the simulation mode, which allows verification of correctness of encoded gates.
The final output from the API is the UDP (user datagram protocol) packet, which will be sent out over Gigabit Ethernet to the FPGA hardware.

\section*{Acknowledgment}
This work was supported by the U.S. Department of Energy, Office of Science, Advanced Scientific Computing Research Testbeds for Science program, the High Energy Physics QUANTISED program, and the National Quantum Information Science Research Centers Quantum Systems Accelerator under Contract No. DE-AC02-05CH11231.
The authors would like to thank Larry Doolittle, Qiang Du, Wim Lavrijsen, Thorsten Stezelberger, Anastasiia Butko, Costin Iancu from Lawrence Berkeley National Laboratory, and Akel Hashim, Jean-Loup Ville, Jie Luo, Brian Marinelli, John Mark Kreikebaum, Yosep Kim, Larry Chen, Will Livingston, Noah Stevenson, Gerwin Koolstra, Marie Lu from University of California, Berkeley, and Quantum Benchmark, Inc. for their support.

\end{document}